\documentclass[galley]{mn2e}
\usepackage{psfig}

\title[Boundary Corrections in Fractal Analysis of Galaxy Surveys]
{Boundary Corrections in Fractal Analysis of Galaxy Surveys}
\author[Jun Pan and Peter Coles]{Jun Pan\thanks{e-mail:
ppxjp@nottingham.ac.uk} and Peter Coles\thanks{email:
Peter.Coles@Nottingham.ac.uk} \\School of Physics \& Astronomy,
University of Nottingham, University Park, Nottingham NG7 2RD, UK
}

\begin{document}
\maketitle

\begin{abstract}
The analysis of redshift surveys with fractal tools requires one
to apply some form of statistical correction for galaxies lying
near the geometric boundary of the sample. In this paper we
compare three different methods of performing such a correction
upon estimates of the correlation integral in order to assess the
extent to which  estimates may be biased by boundary terms. We
apply the corrections illustrative examples, including a simple
fractal set (L\'{e}vy flight), a random $\beta$-model, and a
subset of the CfA2 Southern Cap survey. This study shows that the
new ``angular'' correction method we present is more generally
applicable than the other methods used to date: the conventional
``capacity'' correction imposes a bias towards homogeneity, and
the ``deflation'' method discards large-scale information, and
consequently reduces the statistical usefulness of data sets. The
``angular'' correction method is effective at recovering
true fractal dimensions, although the extent to which boundary
corrections are important depends on the form of fractal
distribution assumed as well as the details of the survey
geometry. We also show that the CfA2 Southern sample does not show
any real evidence of a transition to homogeneity. We then revisit
the IRAS PSCz survey and ``mock'' PSCz catalogues made using
N-body simulations of two different cosmologies. The results we
obtain from the PSCz survey are not significantly affected by the
form of boundary correction used, confirming that the transition
from fractal to homogeneous behaviour reported by Pan \& Coles
(2000) is real.
\end{abstract}


\section{Introduction}

There has been considerable controversy over a long period about
whether the distribution of galaxies in the Universe is
homogeneous on large scales, or whether structure might persist
indefinitely in the manner of fractal (e.g. Coleman \& Pietronero
1992, hereafter CP92; Peebles 1993). This debate revolves around
the validity of the Cosmological Principle on scales up to the
present observational depth of galaxy surveys. The assumption of
homogeneity plays such an important role in cosmology that it is
important to establish its validity in the most rigorous possible
manner using the most appropriate statistical tools. Instead of
the usual two-point correlation function, $\xi(r)$, which assumes
homogeneity at the outset, these studies usually exploit a
function known as the conditional density, which does not assume
that the global mean density of points $\langle n \rangle$ is determined by the
sample in question, or even that it exists at all. The global mean
density for a fractal set is not a useful concept in any case
(Coleman et al 1988; CP92). For non-fractal set for which the mean
density is well-defined, the conditional density $\Gamma(r)$ is
related to the usual two-point correlation function via
\begin{equation}
\frac{\Gamma(r)}{\langle n \rangle}=1+\xi(r)=g(r).
\end{equation}
One of the consequences of a fractal distribution is that the
quantity $g(r)$ has a power-law dependency on scale:
\begin{equation}
g(r) \propto r^{D-3},
\end{equation}
where $D$ is (loosely) called the fractal dimension; see later for
more rigorous definitions. A simple fractal will have a constant
value of $D$, whereas a structure that tends towards uniformity
will have $D\rightarrow 3$ at some scale.

Some authors have claimed to have found a transition from
quasi-fractal behaviour on small scales to a homogeneous behaviour
on large scales, with a crossover depth around $30h^{-1}$ Mpc ($h$
is Hubble's constant in units of 100 km s$^{-1}$ Mpc$^{-1}$).
Examples of data sets analysed this way include the Perseus-Pisces
survey (Guzzo et al. 1991), the CfA slice (Lemson \& Sanders
1991), the ESO Slice Project (Scaramella et al. 1998) and the Las
Campanas redshift Survey (Kurokawa, Morikawa \& Mouri 2001). There
are other papers in support of homogeneity on large scale, but
advocating a different scale at which the transition occurs
(Hatton 1999; Bharadwaj, Gupta \& Seshadri 1999).

On the other hand, advocates of a purely fractal Universe argue
that, up to the presently-observed scales, there is no indication
of such a transition at all and that the apparent crossover at $30
h^{-1}$ Mpc claim is spurious. Possible causes of this spurious
detection are: the use of inappropriate statistical tools, i.e.
the two-point correlation $\xi(r)$; errors resulting from
uncertainties in the K-correction; misleading boundary
corrections; and so on (e.g. CP92; Sylos Labini et al 1996; Joyce
et al 1999; Best 2000). In this paper we address the last of these
issues, the one which is most often flagged as a possible
mechanism for forcing a fractal distribution to display a false
signature of homogeneity. Previous papers have examined the
behaviour of the conditional density under various forms of
boundary correction, with the conclusion is that  $g(r)$ and
consequently the scale of the crossover are almost independent of
such boundary corrections (Lemson \& Sanders 1991; Provenzale,
Guzzo \& Murante 1994). These arguments are largely based on their
analysis of CfA and mock data sets.

In this work we examine the effect on the correlation integral
(CI), mathematically the integral form of the conditional density
mentioned above (Grassberger \& Procaccia 1983; Borgani
1995). Boundary corrections appear more explicitly in the CI
approach than with the conditional density, which has led to a
criticism of this approach (CP92; Marcelo \& Alexandre 1998; Sylos
Labini et al 1996). However, the only method which is free from
any possible effect of boundaries is one in which no correction is
used at all (e.g. Pietronero, Montuori \& Labini 1997). This
inevitably reduces the effective depth of the survey and reduces
the number of galaxies. This, in turn, makes it harder to see
any transition to homogeneity and reduces the statistical
confidence of any analysis method. The motivation for this work is
to find a recipe for dealing with boundary effects that offers a
reasonable compromise between full use of the catalogues and the
possible biases induced by boundary problems, as guide for future
analysis of the forthcoming catalogues, such as the 2DF galaxy
redshift survey (Colless et al. 2001). It is also important to
establish the robustness of the results we obtained in a previous
paper for the PSCz (Pan \& Coles 2000, hereafter PC) in the light
of this comparison.

We will begin in Section 2 with brief description of the CI and
discuss the role of boundary corrections in Section 3. In Section
4 we apply the different methods to some illustrative examples. We
then, in Section 5, we revisit the PSCz survey studied by PC,
alongside  mock catalogues made from N-body simulations. The
conclusions and a discussion are in Section 6.

\section{The Correlation Integral}

The measure we use for fractal dimension estimation is constructed
from the partition function,
\begin{equation}
Z(q,r)=\frac{1}{N}\sum_{i=1}^{N} p_i(r)^{q-1}\propto r^{\tau (q)},
\end{equation}
with $p(i)=n_i(r)/N$, where $n_i(r)$ is the count of objects in
the cell of radius $r$ centered upon an object labelled by $i$
(which is not included in the  count). For each value of $q$ in
equation (3) correponding to relevant moment of the cell-count,
one can have a different scaling exponent of the set $\tau(q)$,
which induces the so-called Renyi dimensions:
\begin{equation}
D_q=\frac{\tau(q)}{q-1}
\end{equation}
forming the spectrum of fractal dimensions for a fractal measure
on the sample. The terminology applied to a set in which the $D_q$
are functions of $q$ is a {\em multifractal}. The special case
$D_1$ for $q=1$ cannot be obtained from equation (3) but should be
derived from
\begin{equation}
S(r)= \frac{1}{N}\sum_{i=1}^{N}\log p_i(r) \propto r^{D_1},
\end{equation}
where $S(r)$ is the partition {\em entropy} of the measure on the
sample set; $D_1$ is consequently termed the information dimension
(Fedar 1988). The special case $q=2$ leads to the definition of
$D_2$, described in equation (3). This is the most important
exponent in this context and is generally called the correlation
dimension. As stated above, it is related to the usual two-point
correlation function $\xi(r)$ for a sample displaying large-scale
homogeneity (Peebles 1980). If the mean number of neighbours
around a given point is $\langle n \rangle$ then
\begin{equation}
\langle n \rangle=4\pi\bar{n} \int_{0}^{r} [1+\xi(s)]s^2 ds.
\end{equation}
In this case $\langle n \rangle  \sim r^{\alpha}$ means
$\alpha=D_2$. A homogeneous distribution has $D_2=3$, whereas a
power-law in $1+\xi(r)=g(r) \sim r^{-\gamma}$ yields
$D_2=3-\gamma$.

To account for edge effects and the selection function, we have to
weight local count around $i$th object according to
\begin{equation}
n_i(r)=\frac{1}{f_i(r)} \sum_{j=1}^{N} \frac{\Psi(|\textbf{r}_j -
\textbf{r}_i| - r)}{\phi(r_j)},
\end{equation} and
\begin{equation}
\Psi(x)=\left\{\begin{array}{cc} 1, & x \leq 0\\ 0, & x> 0
\end{array}\right..
\end{equation}
For magnitude (or flux) limited sample $\phi(r)$ is exactly the
luminuosity selection function, while $\phi(r)=1$ for volume
limited sample. It is in the weighting factor $f_i(r)$ that the
question of appropriate boundary corrections is most important.

\section{Boundary Correction Methods}
In the literature, there are two standard ways of handling
boundary corrections in this type of analysis. The obvious one is
the {\em capacity correction} which has been used in a series of
papers analysing galaxy catalogues (Mart\'{\i}nez \& Coles 1994;
Pan \& Coles 2000) and cluster catalogues (Borgani \&
Mart\'{\i}nez et al. 1994). In this prescription, counts for those
cells near the boundary are weighted by a factor $f_i(r)$
determined by the capacity (for 3-dimensions this is the volume)
of the cell with radius $r$ included within the sample space. This
is probably the most natural idea how to perform a boundary
correction,  but it may give rise to artificial homogeneity of the
sample, as pointed out for example by CP92. Consider the example
of a ball of radius $r$ with points inside it distributed
according to a density law $n\propto r^{D}$ with $D<3$. If we
extend the cell count around the ball's centre to scale $R>r$
using the capacity correction, we will be extrapolating the count
to scale $R$ with the law of $n\propto r^{3}$. This criticism is
made forcefully by those advocating the purely fractal picture.

A radical way to circumvent this problem is to discard any cells
not completely contained within the sample space; we name this the
{\em deflation method} in this paper. The $N$ in equations (3) and
(5) is then not the total number of galaxies but the number of
those left after elimination. This correction has been used quite
often in estimates based on the conditional density (e.g. CP92).
The largest scale that can be detected when this method is used is
the radius of the biggest sphere that can be entirely fitted into
the sample space. This is  defined to be the {\em effective sample
radius} $R_s$ and is very much less than the real depth of most
samples, especially `pencil-beam' or 'fan' surveys, or surveys
containing 'holes' due to  variations in completeness of sky
coverage. This correction greatly cut down the effectiveness of
such surveys. Typically, for a sample of depth $\sim 150h^{-1}$
Mpc, the effective depth can be as small as $\sim 20h^{-1}$ Mpc.
This is not an efficient use of the data.

A second crucial shortcoming of this approach is that it is
statistically unreasonable. As the size of a cell $R$ is
increased, the number of cells remaining in the sample decreases
until the effective sample radius $R_s$ is reached. On large
scales, therefore, the correlation integral is usually dominated
by a few cells around galaxies with located at particular places
within the survey geometry. In the sense of statistics this is
rather unfair: the measure $Z(r)$ is averaged over large number of
cells on small scales, while very few are included on large
scales. This reduces the statistical significance of measured
values quite considerably for cells of size similar to the
effective radius.

Note also that some authors (e.g. CP92), have used the conditional
density about the observer rather than around individual galaxies
because of the impossibility of reaching scales larger than $R_s$.
They have thus inferred the validity of a fractal picture up to
about a few hundred Megaparsec. This is highly misleading. First
of all, a power-law around one peculiar point is not proof of
fractal scaling. The behaviour of $g(r)$ around the observer is a
special event and should not be used as a description of general
property of the sample. Furthermore, the $g(r)$ is based on the
points located almost at the sample's centre when $r$ is close to
$R_s$, but in case of $r>R_s$, $g(r)$ is based entirely on the
observing point. These are not measuring the same thing, so any
claimed inhomogeneity based on such an approach begs the question.

Of course any boundary correction has to rely on some assumption
about the distribution of points beyond the sample boundary, and
to some extent this inevitably leads to a bias of some kind. We
have to be aware of what kind of bias this is, and what the
statistical consequences are so that we can interpret our results
in a reasonable way. The discussion above indicates that the two
conventional prescriptions, the capacity and deflation
corrections, do not succeed in these aims. We therefore need to
find new ways to perform a more suitable edge correction. Any
useful  new method should make minimal  assumptions while making
maximal use of the information contained within the sample.

In this spirit, we propose a third correction: the  {\em angular
correction}. This proposal stems from the realisation that the
appropriate measure in equation (3) relates to a scaling law that
depends only on the radius $r$ and which has nothing to do with
the angles. From this point of view, we can construct a correction
relying on the solid angular occupation of the sample relative to
the cell's centre, as illustrated in Figure 1. The correction
requires two steps, the first of which resembles the capacity
correction in fixing the factor $f_i(r)$ and the second is similar
to the deflation method when counting neighbours. For a point $i$
in the sample, we calculate $4\pi [1-f_i(r)]$ as the solid angle
subtended at the point in question by intersection of a sphere of
radius $r$ with the boundary. Let's mark the joint space of the
cones in the sphere opened by these solid angles $w^*$, then
correspondingly during neighbour counting process the points
belonging to $w^*$ are excluded. We can increase $r$ for detection
until either one of the correction factors equals zero.

\begin{figure}
\psfig{file=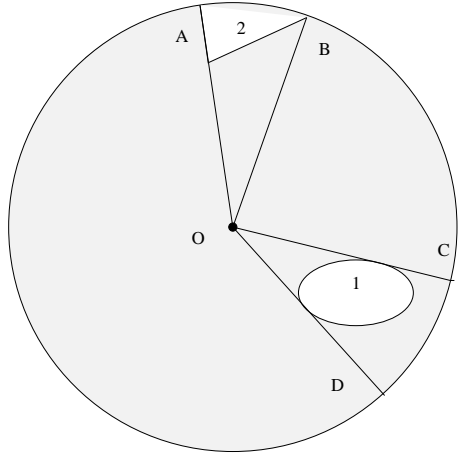,height=6cm,width=6cm,angle=-90} \caption{An
illustration in two dimensions of the angular correction method to
the CI works with a boundary. A cell with radius $r$ is drawn on
point $0$ inside the sample. The shaded region is inside the
sample; two regions not belonging to the sample space are labelled
1 and 2. The angles opened by them are $\theta_{AOB}$ and
$\theta_{COD}$ respectively. The correction factor is then
$f_O(r)=1-(\theta_{AOB}+\theta_{COD})/2\pi$. And when we count the
neighbours, we need to exclude those points in the regions
delimited by $AOB$ and $COD$. The application to solid angles in
3D is straightforward as described in the text.}
\end{figure}

The underlying assumption of this method is of statistical
isotropy. The capacity correction relies on homogeneity too,
because it incorporates an extra correction to the radial part of
the cell count. In this correction, the availability of a sample
point depends on its largest distance to a boundary in contrast to
the deflation method where it is the smallest scale that counts.
This does increase significantly the usable depth of the sample.
The maximum scale that can guarantee that all  sample points are
available, which keeps the measure statistically fair, is the
smallest one of the largest distances of these points to the
boundary surface.

\section{Illustrative Examples}
It is hard to show analytically how much these boundary correction
methods reflect the true scaling behavior of samples and which one
is the best order from the menu of edge corrections. We have
instead to turn to numerical tests. Since we are talking about
fractal analysis, it will be worthwhile to consider these methods
in the context of a simple fractal set with known dimension. After
establishing the results of this we can understand their effects
on real samples and catalogues from n-body simulation. The form of
the sample boundary will itself play a role in determining the
importance of boundary corrections. Since these are primarily for
illustration of the possible pitfalls, and we can't in any case
simulate every possible boundary of every possible sample, we will
proceed using sets with relatively simple boundaries.

\subsection{Monofractal set: the random L\'{e}vy flight}
Let us just start with a monofractal sample with dimension $D$,
which is obtained by the simple L\'{e}vy flight (Meakin 1998). In
this case, the $D$ coincides with the correlation dimension $D_2$.
The L\'{e}vy flight is one species of fractal Brownian random walk
with variable step size $X$, such that the probability of $X$
exceeding a particular value $x$ satisfies
\begin{equation}
P(X\geq x/x_0)=\left\{\begin{array}{cc} (x/x_0)^{-D},& x/x_0 > 1\\ 1,
& x/x_0 \leq 1
\end{array}\right.,
\end{equation}
which leads to a fractal point set of dimension $D$ on scales
$X\gg 1$ (Mandelbrot, 1977). The parameter $x_0$ here plays the
role as the minimum step size of the random walk.

We construct a cube-shaped sample of $D=1.2$, roughly that
observed in galaxy clustering. We use a test volume of  $60\times
60 \times 60 $ Mpc, and set up the minimum step size $x_0=0.2$
Mpc. The correlation integrals obtained using the different edge
treatments discussed in Section 3 are shown in Figure 2 and the
 dimensions obtained in different domains are listed in Table 1.
 Because we set the
minimum step size of the L\'{e}vy walk to be $0.2$ Mpc, it is not
surprising that the correlation integrals below $\sim 1$ Mpc,
where they are dominated by discreteness, are quite steep. When
$r\gg x_0$, the capacity correction obviously contaminates the
estimation badly.  Larger and larger values of the dimension are
obtained with increasing scale; this trend is entirely spurious.
The deflation method provides better answer but the local
dimension in this case fluctuates wildly around the true value,
which may arise from the fact that what we measure with equation
(3) on different scales is effectively coming from a different
point set.  The performance of angular correction is promisingly
steady and it accurately recovers the true dimension, which
encourages our conjecture that this one is superior to the others.
\begin{table}
\begin{center}
\begin{tabular}{cccc}\hline
L\'{e}vy Flight  & Capacity & Angular & Deflation \\ \hline
$D_2\begin{array}{c}1\sim10\,{\rm Mpc}\\ >10\,{\rm
Mpc}\end{array}$ & $\begin{array}{c}1.33(1) \\ 1.84(3)\end{array}$
& $\begin{array}{c}1.18(1) \\ 1.19(3) \end{array}$ &
$\begin{array}{c}1.17(4) \\ 1.01(8) \end{array}$ \\ \hline $r_{\rm
max}({\rm Mpc})$ & --- & $41.2$ & $30$ \\ \hline
\end{tabular}
\end{center}
\caption{Correlation dimensions $D_2$ for different scale ranges
of the L\'{e}vy flight generated sample with $D=1.2$. Errors are
from goodness-of-fit considerations only. $r_{\rm max}$ indicates
the largest available scale as discussed in text for angular
correction and deflation method.}
\end{table}

\begin{figure*}
\begin{tabular}{cc}
\psfig{file=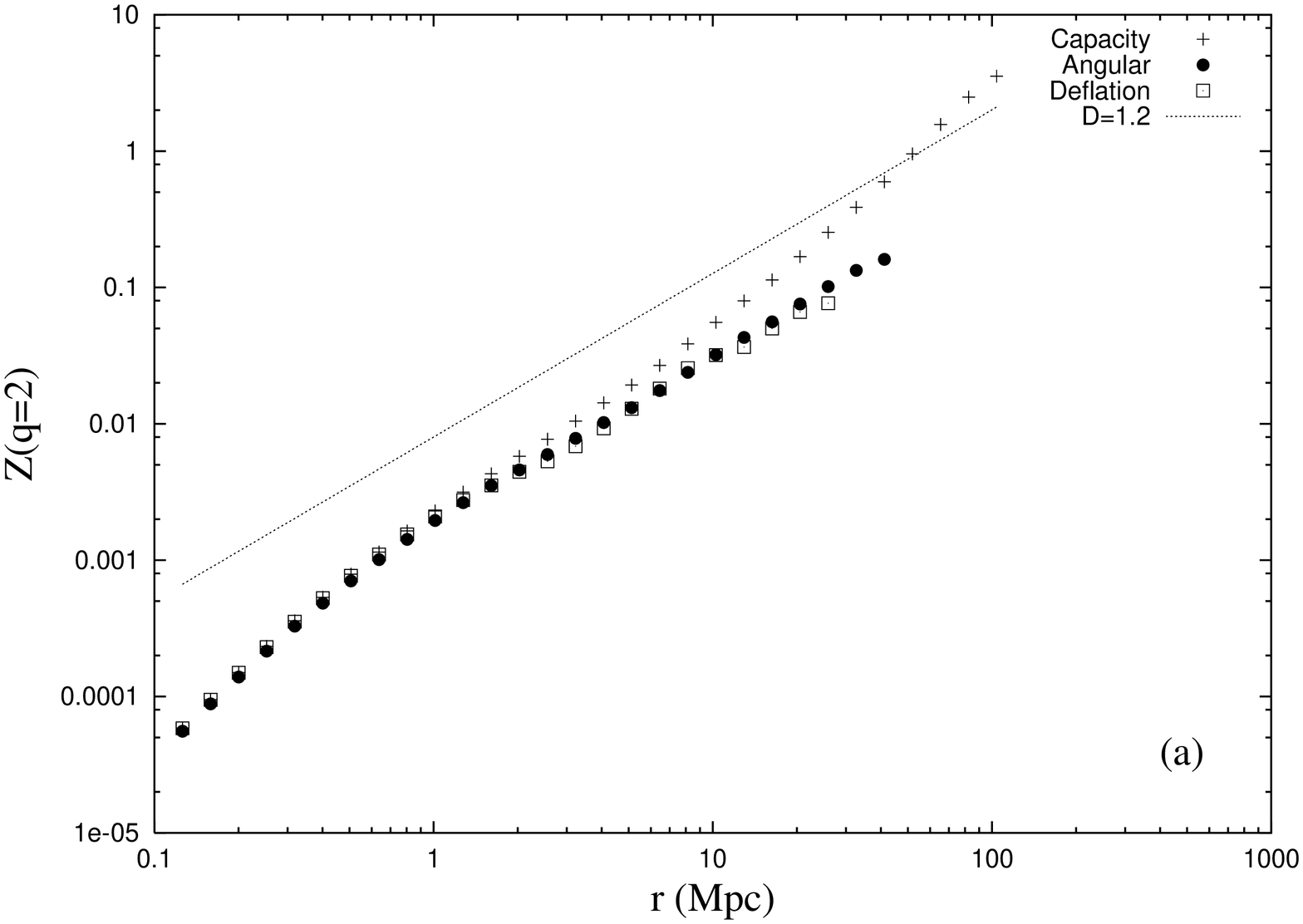,height=8cm,width=7cm,angle=-90}&
\psfig{file=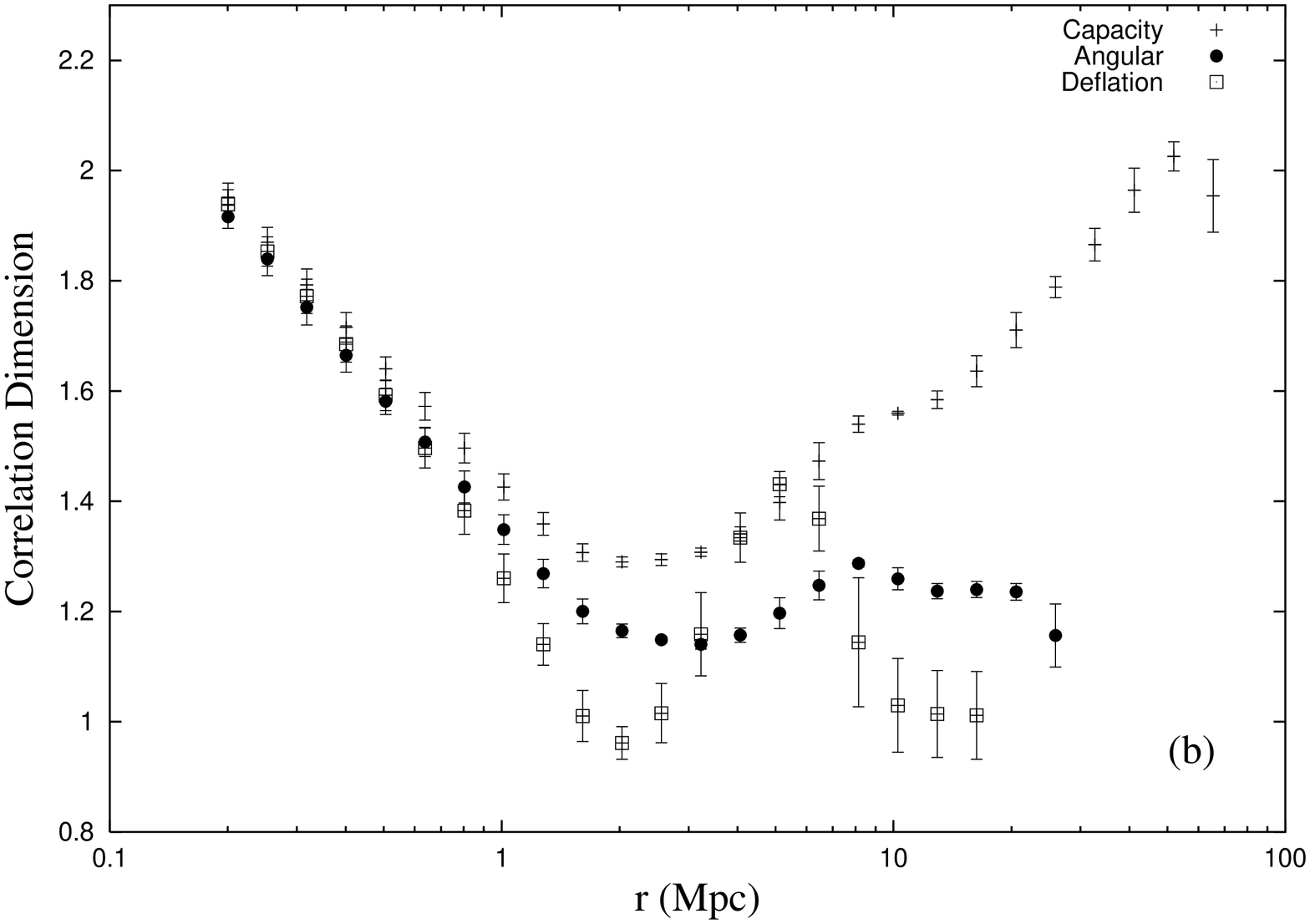,height=8cm,width=7cm,angle=-90}
\end{tabular}
\caption{The correlation integral analysis under three different
boundary correction methods for L\'{e}vy flight generated fractal
point set with dimension $D=1.2$ and minimum step size $x_0=0.2$
Mpc. The sample space is a $60\times 60 \times 60$ Mpc cube. In
this plot and all the plots hereafter, the value of $h$ is set to
be unity. Panel (a) shows the correlation integrals $Z(q=2)$ of
equation (3), the straight line of slope $D=1.2$ is plotted as
reference. Panel (b) displays the local correlation dimensions
against scale by fitting every five consecutive points. The excess
dimension excess below $\sim 1$ Mpc is a resolution effect. It is
very clearly seen that the capacity correction can seriously mask
the true scaling properties. The angular-corrected estimation is
superior in providing quite stable and accurate results to the one
corrected by deflation method. The maximum available scales for
the deflation method and angular correction are $30$ Mpc and $\sim
40$ Mpc respectively.}
\end{figure*}

This example constitutes a rather severe test because each
realisation of the random L\'{e}vy flight is highly anisotropic.
Although the pattern of large-scale structure does display
filaments, they are by no means as exaggerated as this. In the
following we look at a less extreme model.

\subsection{A different example: The  $\beta$-model}
We next examine a simple self-similar cascading $\beta$ model.
Points are generated by a breaking cascade from the parent cube
with size $L_0$ into $M$ smaller but similar cubes with size
$L_1=L_0/n$ (usually $n=2$, thus $M=n^3$). Each cube is then
assigned a survival probability $p$ until the next iteration at
which it stands a chance of breaking again. The final set is the
collection of all the survived points (cubes) after $k$ iterations
(see Castagnoli \& Provenzale 1991). This model is qualitatively
similar to  hierarchical clustering.

Here the survival probability $p$ remains constant for all cubes
and all iterations. The fractal dimension of this simple model is
given by
\begin{equation}
D=\lim_{k\rightarrow \infty}\frac{\log (pM)^k}{\log
n^k}=\frac{\log pM}{\log n}.
\end{equation}

\begin{figure*}
\begin{tabular}{cc}
\psfig{file=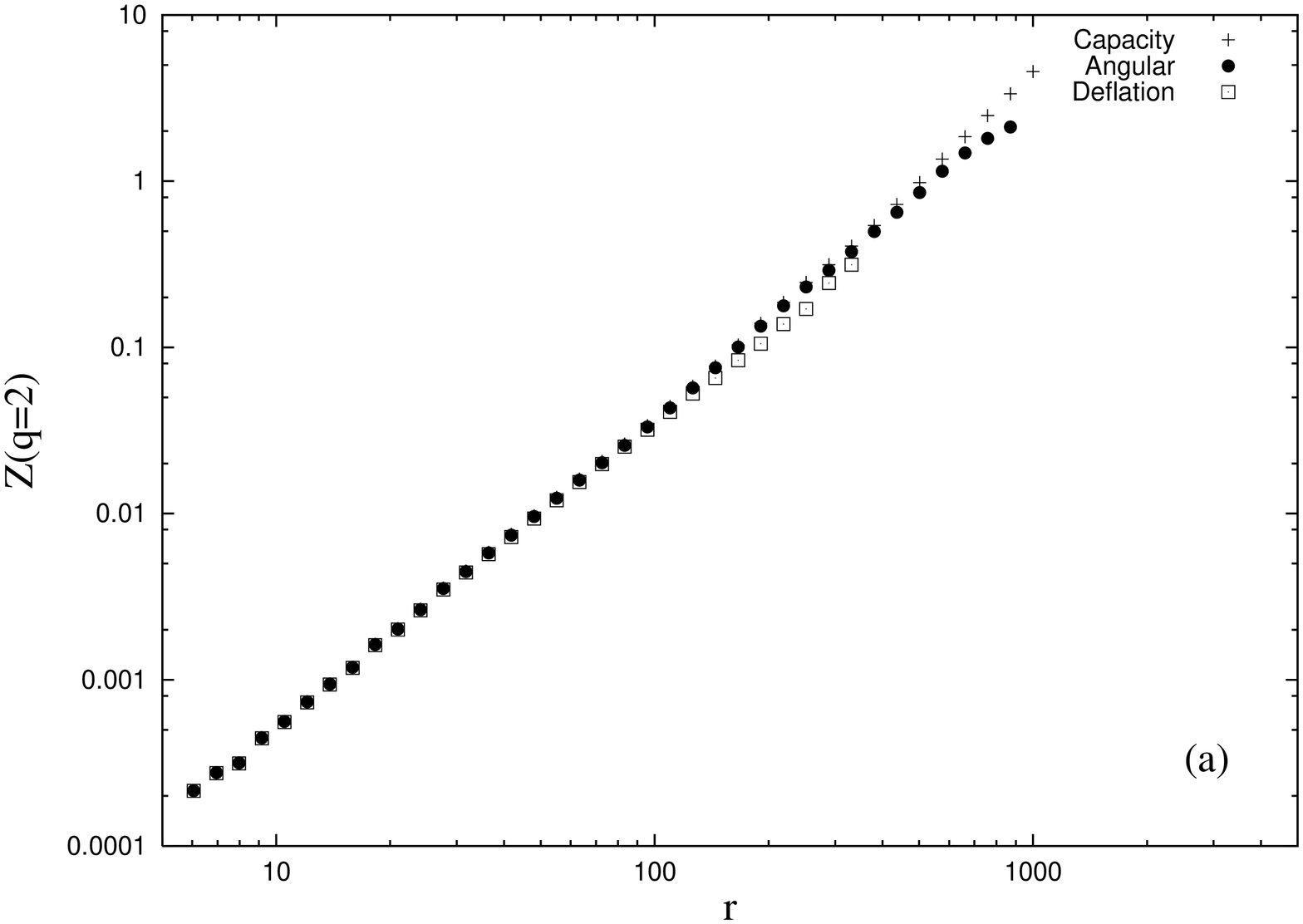,height=8cm,width=7cm,angle=-90}&
\psfig{file=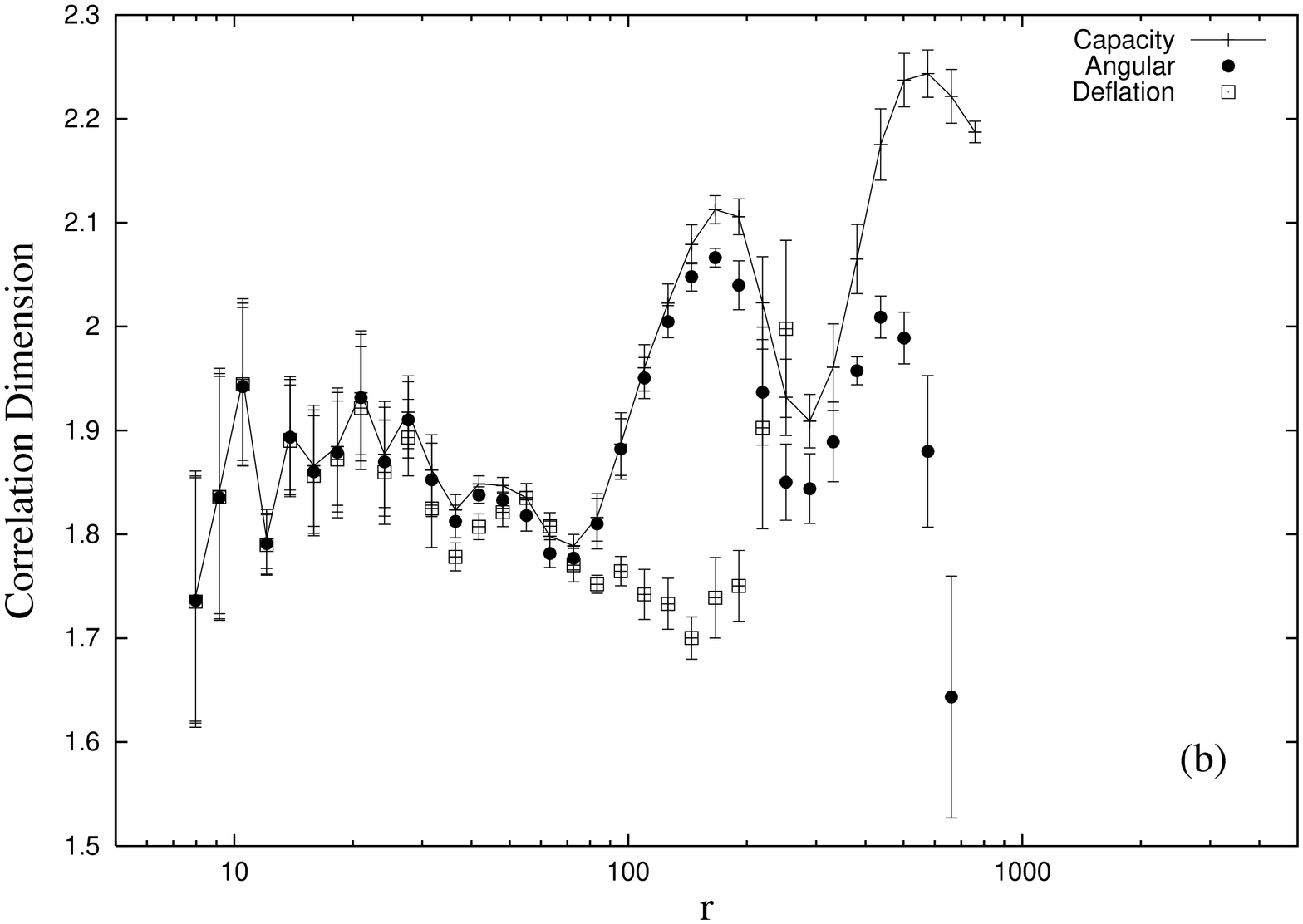,height=8cm,width=7cm,angle=-90}
\end{tabular}
\caption{As Figure 2 (except here arbitrary scale unit is used), but 
for the $\beta$-model. The parameters
chosen are  $L_0=1000$ and $p=0.4352$ corresponding to $D=1.8$.
About 20,000 points are generated for this analysis. The differences
between capacity and angular correction are not so dramatic as the
L\'{e}vy flight example, but we can still identify the systematic 
increase of local dimensions under capacity correction.}
\end{figure*}

\begin{table}
\begin{center}
\begin{tabular}{cccc}\hline
$\beta$ model  & Capacity & Angular & Deflation \\ \hline
$D_2\begin{array}{c}<100\,\\ >100\,\end{array}$ &
$\begin{array}{c}1.86(1) \\ 2.09(1)\end{array}$ &
$\begin{array}{c}1.85(1) \\ 1.92(2) \end{array}$ &
$\begin{array}{c}1.84(1) \\ 1.83(4) \end{array}$ \\ \hline $r_{\rm
max}$ & --- & $870$ & $340$ \\ \hline
\end{tabular}
\end{center}
\caption{Correlation dimensions $D_2$ for different scale ranges
of the $\beta$ model with $D=1.8$. Errors are from goodness-of-fit
considerations only. The quantity $r_{\rm max}$ is the
largest available scale as discussed in text for angular
correction and deflation method.}
\end{table}

We produced the sample using $p=0.4352$ and $n=2$ correponding 
to $D=1.8$. The parent cube which also defines our sample space is of size
$L_0=1000$. About 20,000 points are generated for our analysis.
The results are displayed in Figure 3 \& Table 2. Panel (b) of Figure 3 shows 
that differences in the three methods arise at a scale of $\sim 100$. Below that, 
unlike the highly anisotropic L\'{e}vy flight example, boundary corrections
have little effects. Angular correction agrees with capacity
correction out to a scale of a few hundred, but  
angular correction does not introduce an apparent trend to higher dimensions
on scales larger than $\sim 100$, which capacity correction does.

Although in this case the differences between the three methods
are somewhat less extreme, it is still the case that the angular
correction method is close to the correct answer, while the
capacity correction produces an artificial tendency towards
homogeneity.

\subsection{CfA2-South Survey}

The two toy examples we have displayed illustrate that one should
take care in implementing boundary corrections that may influence
the measured fractal properties of the sample. We now turn to a
real sample, although by now it is of historical importance only.
The example we choose is the well-studied CfA2-South galaxy survey
(Huchra et al. 1999). Previous research has indicated its
dimension $D=1.8\sim 2.0$ up to $\sim 30 h^{-1}$ Mpc approached
with different fractal tools (Joyce, Montuori \& Sylos Labini
1999; Kurokawa, Morikawa \& Mouri 1999). The question we ask is
whether, given the potential dangers we described above, there is
evidence that this survey displays a tendency towards homogeneity?

The sample we study covers $20^h\leq\alpha \leq 4^h$ in right
ascension and $-2.5^{\circ}\leq\delta\leq 90^{\circ}$ in
declination, containing 4390 galaxies in total with magnitude
$m_{B(0)}\leq 15.5$. Following Park et al. (1994), we exclude
areas where there is significant interstellar extinction from our
Galaxy: $20^h\leq\alpha\leq 21^h$; $3^h\leq\alpha\leq 4^h$;
$21^h\leq \alpha\leq 2^h$ and $b>-25^{\circ}$; $2^h \leq\alpha\leq
3^h$ and $b>-45^{\circ}$. Here $b$ is the Galactic latitude. The
distances are computed from the redshift $z$ using the Mattig
formula,
\begin{equation}
r=\frac{c}{H_0 {q_0}^2 (1+z)}[q_0 z + (q_0 -1)\cdot(\sqrt{2 q_0 z +1}-1)] \, ,
\end{equation}
with $H_0=100h^{-1}$ km s$^{-1}$ Mpc$^{-1}$  and $q_0=0.5$.

We construct a volume-limited sample with magnitude threshold
$M_{B(0)}=-18.46$. The absolute magnitude is obtaineed from
\begin{equation}
M_{B(0)}=m_{B(0)}-5\log[r(1+z)]-25-Kz\,,
\end{equation}
where the $K$-correction factor $K$ here is taken to be $3$ (Park
et al 1994). The sample thus has 766 galaxies, and its depth is
$60h^{-1}$ Mpc corresponding to a redshift $z=0.02$.

\begin{figure*}
\begin{tabular}{cc}
\psfig{file=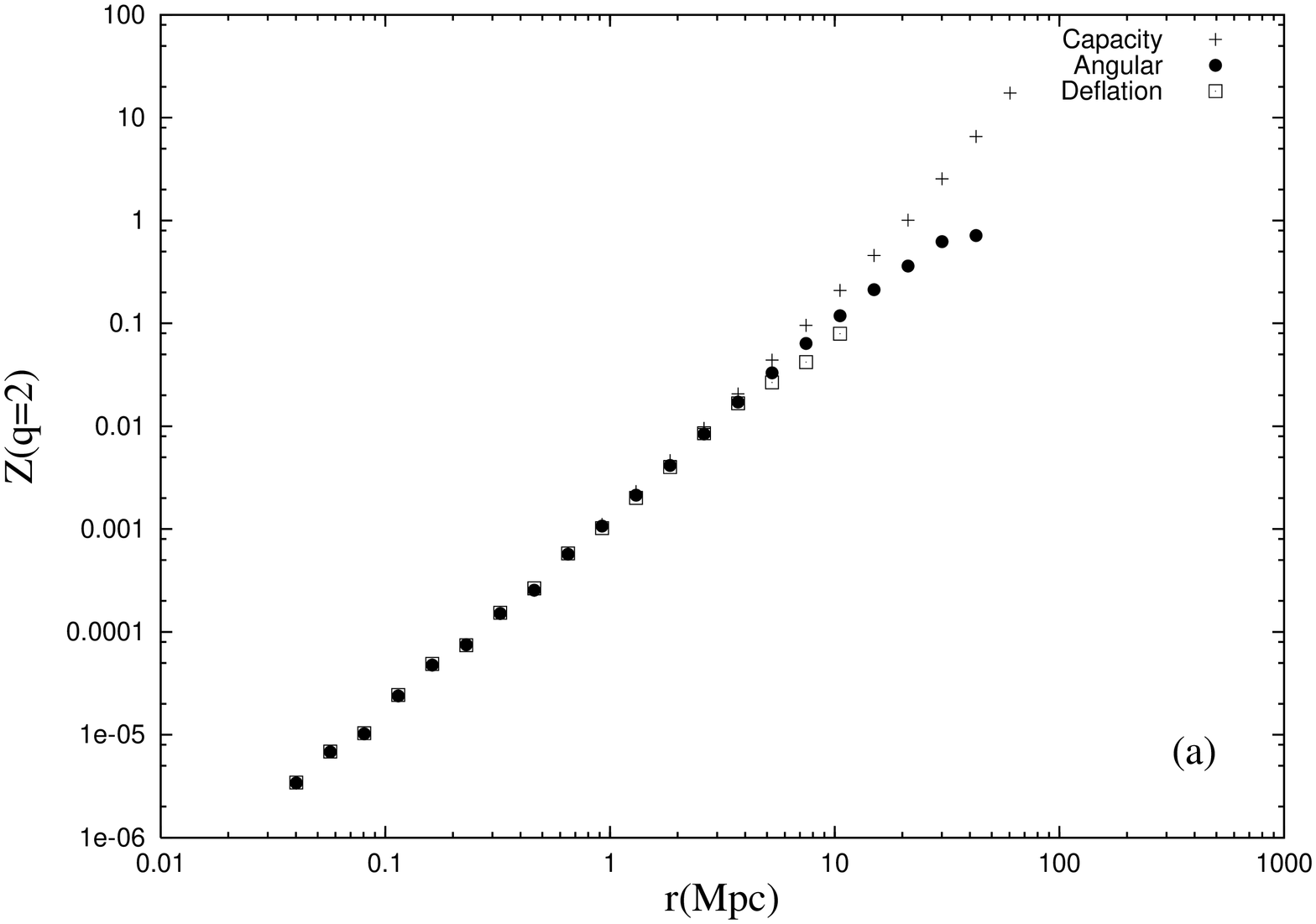,height=8cm,width=7cm,angle=-90}&
\psfig{file=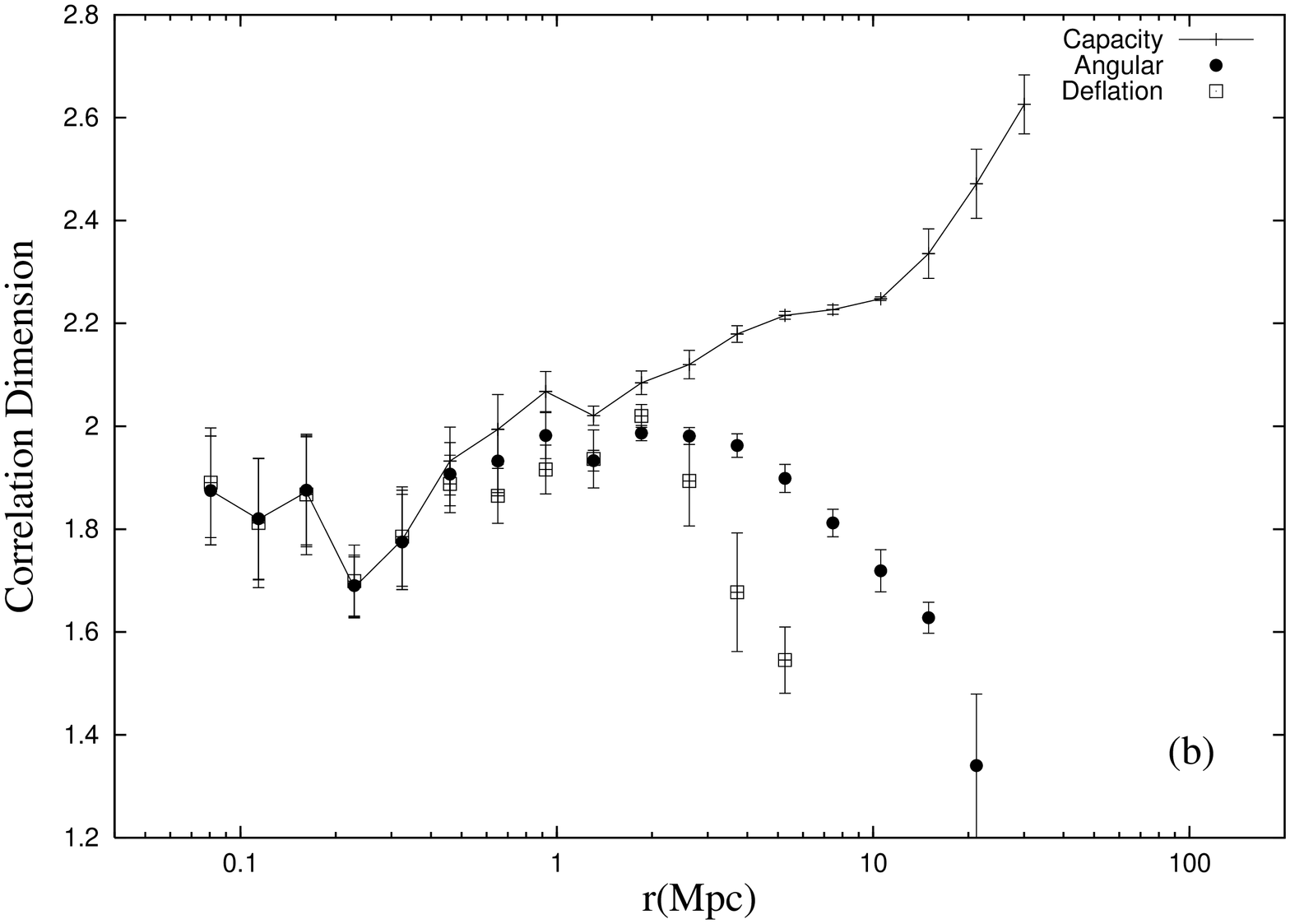,height=8cm,width=7cm,angle=-90}
\end{tabular}
\caption{Correlation integrals of CfA2-South volume-limited sample
from different boundary corrections. The sample size is $60h^{-1}$
Mpc, the minimum and maximum separation between galaxies are
$0.03h^{-1}$ Mpc  and $76h^{-1}$ Mpc. As in Fig. 2, panel (a)
shows the measure $Z$; and panel (b) gives the local $D_2$, in
which the points from the capacity correction are connected with a
line which shows larger dimensions than the other two different
methods, and also the results from Joyce et al (1999). }
\end{figure*}

\begin{table}
\begin{center}
\begin{tabular}{cccc} \hline
CfA2-South  & Capacity & Angular & Deflation \\ \hline $D_2$ &
$2.07(3)^*$ & $1.84(2)$ & $1.83(2)$\\ \hline $r_{\rm max}(\rm
Mpc)$ &
--- & $\sim43$ & $\sim11$ \\ \hline
\end{tabular}\\
$^*$ If we fit it above $10 h^{-1}$ Mpc, $D_2=2.54\pm 0.03$.
\end{center}
\caption{Correlation analysis of CfA2-South sample. $D_2$ is
gained via linear regression of $\log(Z)$ vs $\log(r)$ in full
detected scale range. As before, $r_{\rm max}$ is the maximum
scale.}
\end{table}

Our calculations are illustrated in Figure 4 and Table 3. The
results are quite consistent with the analysis of Joyce et al
(1999). They demonstrate the fractal nature of the CfA2-South
galaxy survey sample, with $D_2\sim 1.8$ but with a scale
extending to $\sim 40h^{-1}$ Mpc with our new angular correction
rather than the largest $R_s\approx 30h^{-1}$ Mpc obtained from
the  volume-limited sample VL205 (with limiting absolute magnitude
$M_{B(0)}=-20.5$) using the deflation method.

On small scales, less than $\sim 10h^{-1}$ Mpc, there is little
difference among the measures obtained after correction by
different methods. This can be easily understood because in this
case only a very small number of points needs correction. On
larger scales it becomes apparent that the capacity correction
produces a trend leading to a homogeneous dimension, similar to
the result of L\'{e}vy flight. Again, the deflation method and
angular correction present behavior consistent with each other.
Although the capacity correction does not deviate seriously from
the dimension obtained by fitting over the whole range of scales
in this case, it definitely disguises the behaviour of the sample
with an inclination towards homogeneity (Table 3). It is not clear
what is happening with the deflation and angular corrected
measures on large scales, but similar fluctuations have already
been shown in the L\'{e}vy flight simulations (panel (b) in Figure
2). At least there is no tendency to introduce artificial
homogeneity, and we see that angular correction demonstrates a
more stable measure, i.e. with smaller fluctuations  than the
deflation method.

\subsection{Comments}
At this point we can already form a couple of preliminary
conclusions about this case and that of the toy fractal sets.
First, it is clear that the capacity correction is, in general,
{\em not} the most appropriate available for fractal analysis. The
improper imposition of boundary corrections can substantially
confuse the issue of whether a given sample reaches homogeneity or
not. On the other hand, our new angular correction behaves well in
recovering the true scaling law, with less fluctuations, more
effective use of the sample, and a higher level of reliability
than the deflation method.

\section{The PSCz Survey Revisited}
In a previous paper (PC), we analyzed the IRAS infrared galaxy
redshift catalogue known as the  PSCz survey, with was analyzed
with correlation integral using the conventional capacity
correction. In that paper it is claimed that the correlation
dimension $D_2$ above $30h^{-1}$ Mpc is very close to 3 but it
still has a  value as small as $2.16$ under $10h^{-1}$ Mpc. This
argues for a transition from fractal to homogeneity within the
range $\sim 10$ to $\sim 30h^{-1}$ Mpc. As we have seen in Section
4, this scale range is very close to the range where the capacity
correction begins to effect the true scaling properties of fractal
distributions. The question then arises whether the transition
phenomenon observed by PC in the PSCz may only be an artifact of
the use, in that paper, of the capacity correction. Does the
transition to homogenity still appear if we use different boundary
corrections?

\subsection{Application to the PSCz}
A problem of dealing with real samples like this can be their
complicated geometrical shape. The PSCz sample, for instance, has
troublesome irregular masks (Saunders et al 2000). In particular,
the blank strip running across the sky along a longitude line
makes it difficult to apply the angular correction  directly to
the CI measure. Given this difficulty, the correction factors from
capacity and angular corrections are estimated via Monte Carlo
simulation. We generate sufficient uniformly-distributed points
within each cell at a specific scale and simply count how many lie
within the sample space. However,  the angular correction is still
fairly tricky even within this approach. In this case the Monte
Carlo simulation does not only involve approximating the
edge-correction factors. Neighbour counting is also problematic if
the sample space does not have a simply-connected convex geometry,
such as if there are holes or cuts in it. These factors
dramatically increase the time of computation needed to apply the
boundary corrections.

In order to keep the calculations within a manageable bound, we
therefore used a subsample of the data used in PC, located in the
north galactic hemisphere  for the purpose of comparing different
boundary correction methods. The other selection criteria are kept
as in PC except for the following, latitude $b>30^{\circ}$,
longitude $l<120^{\circ}$ or $l>300^{\circ}$ and the distance to
the galctic equator plane $r\sin b >10 h^{-1}$ Mpc. This subsample
contains 1941 galaxies and has convex boundary surface.

\begin{figure*}
\begin{tabular}{cc}
\psfig{file=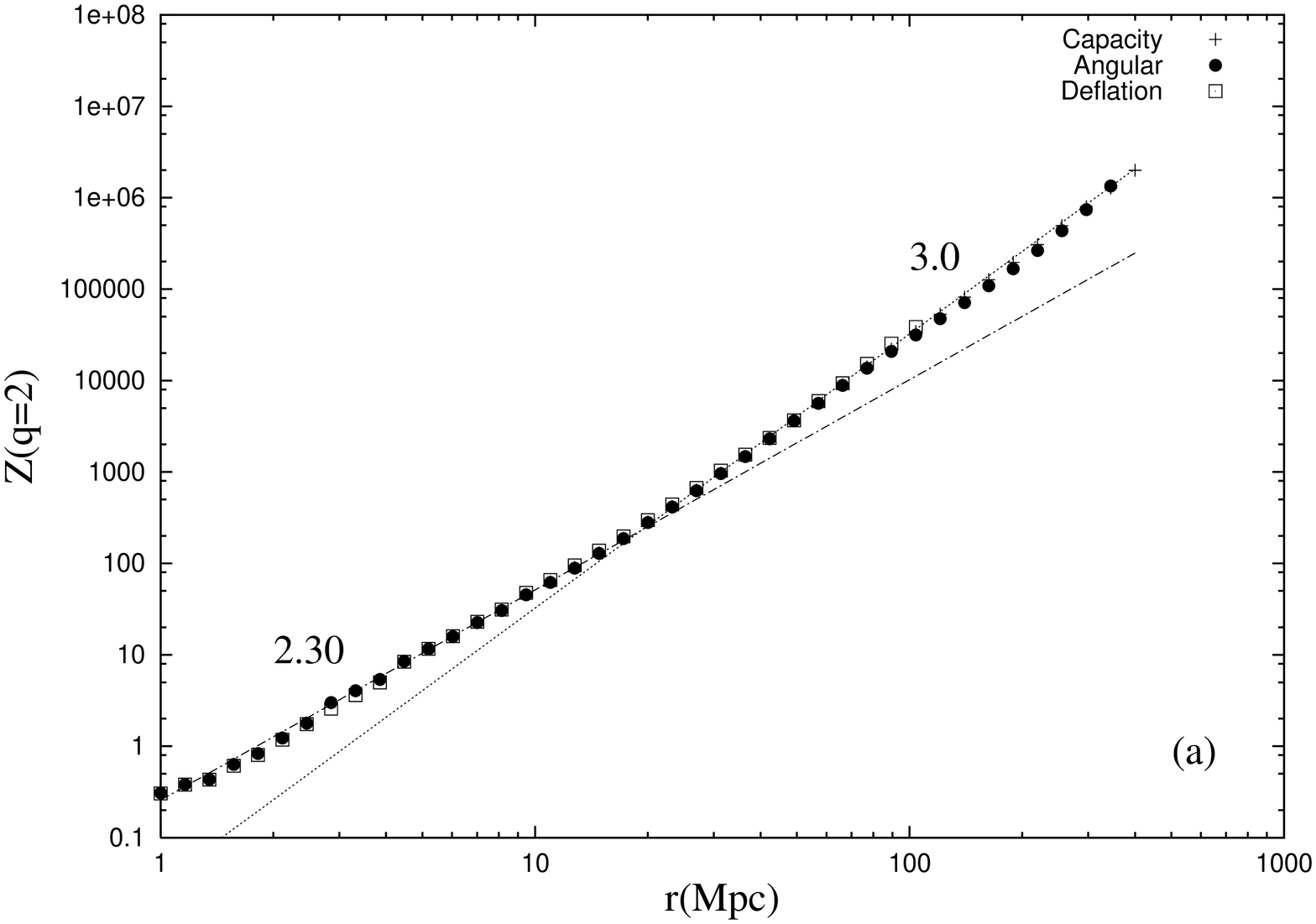,height=80mm,width=70mm,angle=-90}&
\psfig{file=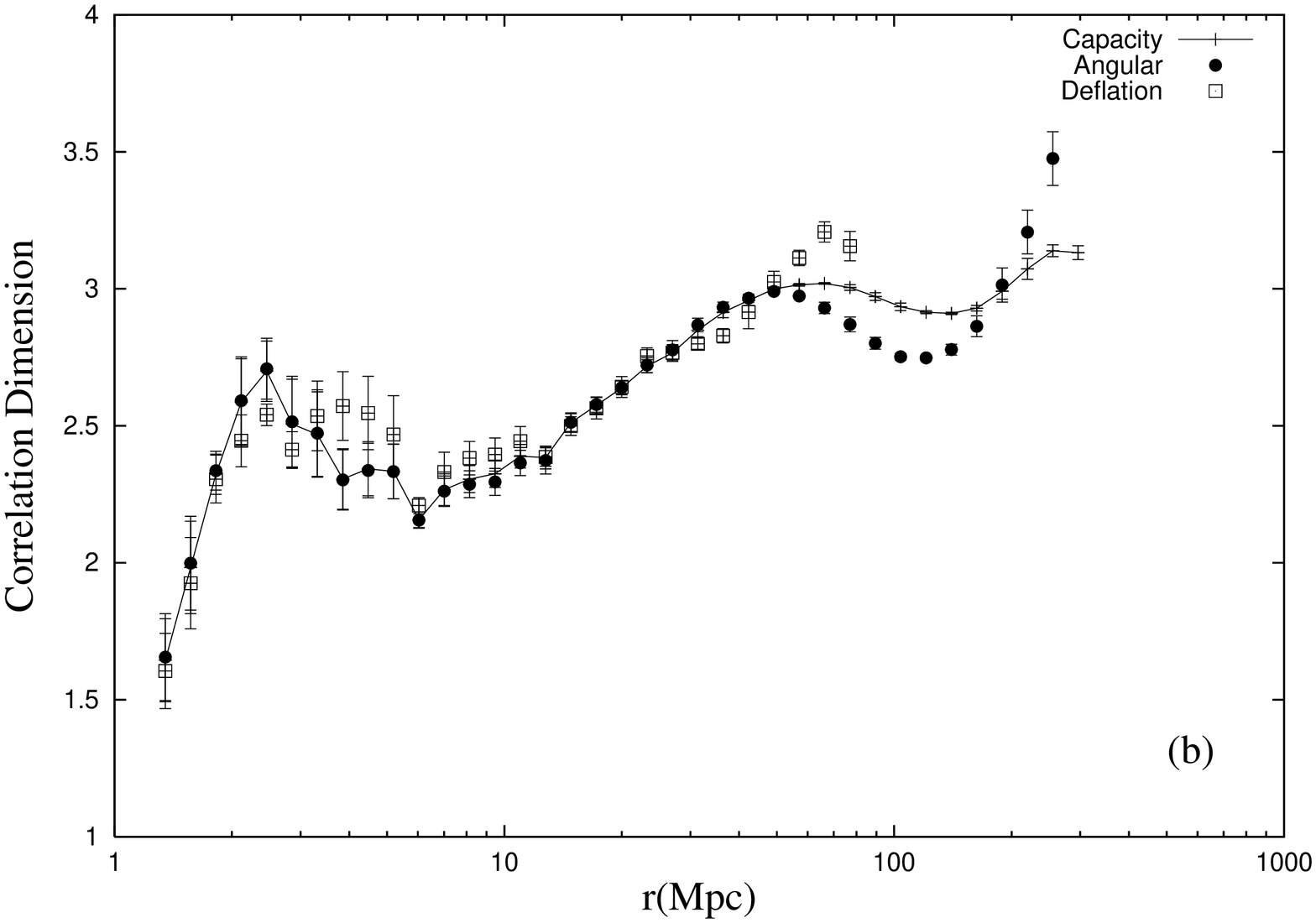,height=80mm,width=70mm,angle=-90}
\end{tabular}
\caption{Application of the method to the PSCz subsample. The left
panel shows $Z(q=2)$ under the different boundary corrections. Two
reference lines with $D_2=2.30$ and $D_2=3.0$ are fitted in
regions $r<10$ Mpc and $r>30$ Mpc. The local  $D_2$ are obtained
by fitting 5 consecutive points  in the right panel; the results
from the  capacity correction are connected with a line. }
\end{figure*}

\begin{table}
\begin{center}
\begin{tabular}{cccc}\hline
PSCz & Capacity & Angular& Deflation \\ \hline $D_2$
$\begin{array}{c}<10\,{\rm Mpc} \\ 10\sim30\,{\rm Mpc}\\>30\,{\rm
Mpc}\end{array}$ & $\begin{array}{c}2.31(4)\\2.57(3)\\2.99(1)
\end{array}$ & $\begin{array}{c}2.32(4)\\2.57(3)\\2.94(3)
\end{array}$ & $\begin{array}{c}2.34(4)\\2.57(3)\\3.06(4)
\end{array}$\\ \hline $r_{\rm max}({\rm Mpc})$ &---&$\sim 344$ & $\sim 104$
\\ \hline
\end{tabular}
\end{center}
\caption{The correlation dimension$D_2$ for different scales of
the PSCz subsample under three boundary correction methods and the
correponding largest scale explorable. Errors are from
goodness-of-fit considerations only.}
\end{table}

The results are shown in Figure 5. The mean separation between
galaxies in this subsample is $\sim 2.6h^{-1}$ Mpc. Below this
scale there is no proper scaling law. The correlation dimensions
obtained for different zones above it are listed in Table 4. It is
a surprise that none of the boundary correction methods affects
either the measure or the estimation of dimensions significantly.
On small scales, it is expected that only a very small fraction of
galaxies is near the sample edge so that the boundary correction
does not have any important influence. On large scales we have
already known from analysis of the previous two samples that the
capacity correction can bias the measure seriously, but this does
not seem to produce a big effect here. The results corrected in
different ways are almost the same until the scale reaches $\sim
30h^{-1}$ Mpc, where the local dimension curves begin to diverge.
Even at this separation the results appear to fluctuate only
around the the homogeneous dimension $D_2=3$. This can only be
explained as the intrinsic distribution is statistically
homogeneous above $30h^{-1}$ Mpc; any boundary correction will
then be unable to deflect the result away from the expected
$D_2=3$ except for chance statistical fluctuations.

The value of $D_2$ for $r<10 h^{-1}$ Mpc is larger than that given
in PC, $D_2=2.15$, but close to the value deduced from QDOT
(Mart\'{\i}nez \& Coles 1994), $D_2=2.25$. The difference between
these three samples is the number of galaxies. There are 1561
galaxies in the QDOT sample, similar to the sample studied here.
This discrepancy may be an  example of the finite size effect
discussed by Sylos Labini et al (1996), or it could be due to
redshift-space distortions, but the transition from fractal to
homogeneity is still very clear. In order to confirm this, we
applied the deflation method to the original sample used in PC;
this gives $D_2(r>30h^{-1} \,{\rm Mpc})=2.98\pm 0.01$.

Now we can say with confidence the conclusion in PC that there is
transition from fractal to homogeneous on scales around $30h^{-1}$
Mpc in this PSCz survey even though PC did not use the safest
method of boundary correction.

\subsection{Mock PSCz catalogues}
Although we have found that analysis of the PSCz sample does not
greatly suffer from boundary correction errors, we still need to
check the validity of our inferences for samples which do display
a mixture of scaling properties, particularly having a gradual
transition to homogeneity. The best choice is to deal with mock
PSCz catalogues from N-body simulation of cosmological models.
This is in any case an interesting exercise, because it may show
us how well the n-body simulations mimic  real clustering when
judged in the light of  a statistical approach that differs from
the more standard correlation functions and power spectra.

We have studied two different mock PSCz catalogues. The catalogue
from SCDM simulates universe with $\Omega_M=1$ and shape parameter
$\Gamma=0.5$ normalized with COBE data. The other one from
$\Lambda$CDM set up $\Omega_{\Lambda}=0.7$, $\Omega_M=0.3$ and
$\Gamma=0.25$. The details of simulation are described in Cole et
al (1998). Both samples are $170$ Mpc deep , include about same
number of points as PSCz and their correlation functions  are also
fitted to agree with PSCz. In order to minimize the effects of
differing geometry, we don't use the the whole data; subsamples
are formed exactly in the way we construct the PSCz subsample
(except for the depth).

The results are reproduced in Figure 6 \& Table 5 and Figure 7 \&
Table 6 repectively for SCDM and $\Lambda$CDM mock data. Neither
simulations shows a sign of significant difference among the three
methods for edge correction. The correlation integral delimits the
transition to homogeneity perfectly.

What is interesting is on small scales, up to $10$ Mpc, both mock
data sets have stronger clustering than the real PSCz survey. The
points numbers of the two mock data samples are quite close to the
real sample, so this over-clustering should not be interpreted as
finite size or discretness. More likely it is due to the
difficulty of finding a prescription for identifying galaxies in
simulations that follow only the evolution of dark matter. One
aspect of this problem is the well-known one that standard CDM
models have too much clustering power on small scales, requiring
some form of anti-bias to be invoked to explain the observations.

\begin{table}
\begin{center}
\begin{tabular}{|c|ccc|}\hline
SCDM & Capacity & Angular& Deflation \\ \hline $D_2$
$\begin{array}{c}<5{\rm Mpc} \\ 5\sim30{\rm Mpc}\\>30{\rm
Mpc}\end{array}$ & $\begin{array}{c}0.97(2)\\2.23(6)\\2.91(2)
\end{array}$ & $\begin{array}{c}0.96(2)\\2.23(6)\\2.88(3)
\end{array}$ & $\begin{array}{c}0.98(3)\\2.02(6)\\2.79(6)
\end{array}$\\ \hline $r_{\rm max}({\rm Mpc})$ &---&$\sim 132$ & $\sim 54$
\\ \hline
\end{tabular}
\end{center}
\caption{Correlation dimensions $D_2$ and for different scales of
the SCDM mock PSCz subsample and $r_{\rm max}$ for angular
correction and deflation method. Errors are from goodness-of-fit
considerations only.}
\end{table}

\begin{figure*}
\begin{tabular}{cc}
\psfig{file=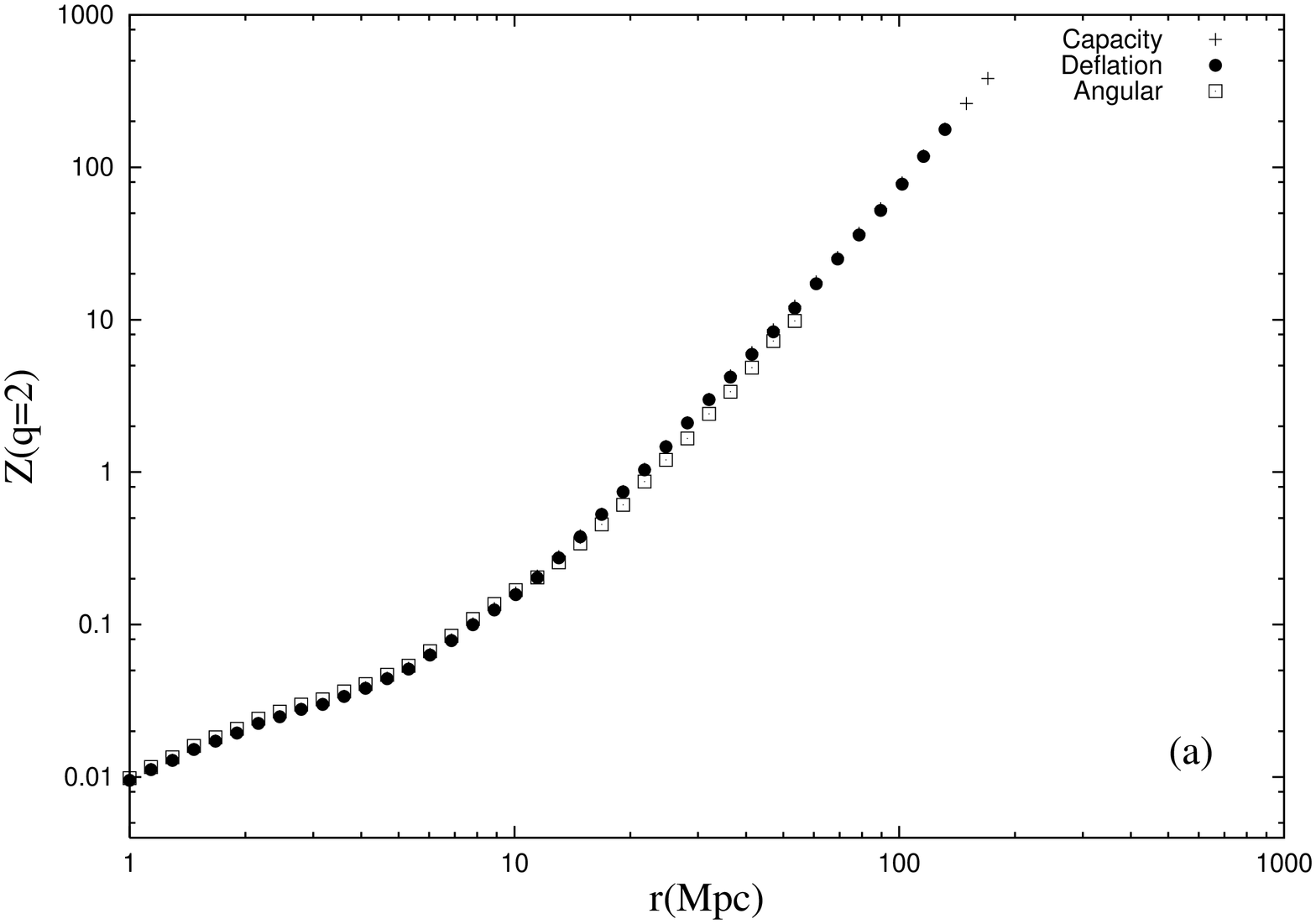,height=8cm,width=7cm,angle=-90}&
\psfig{file=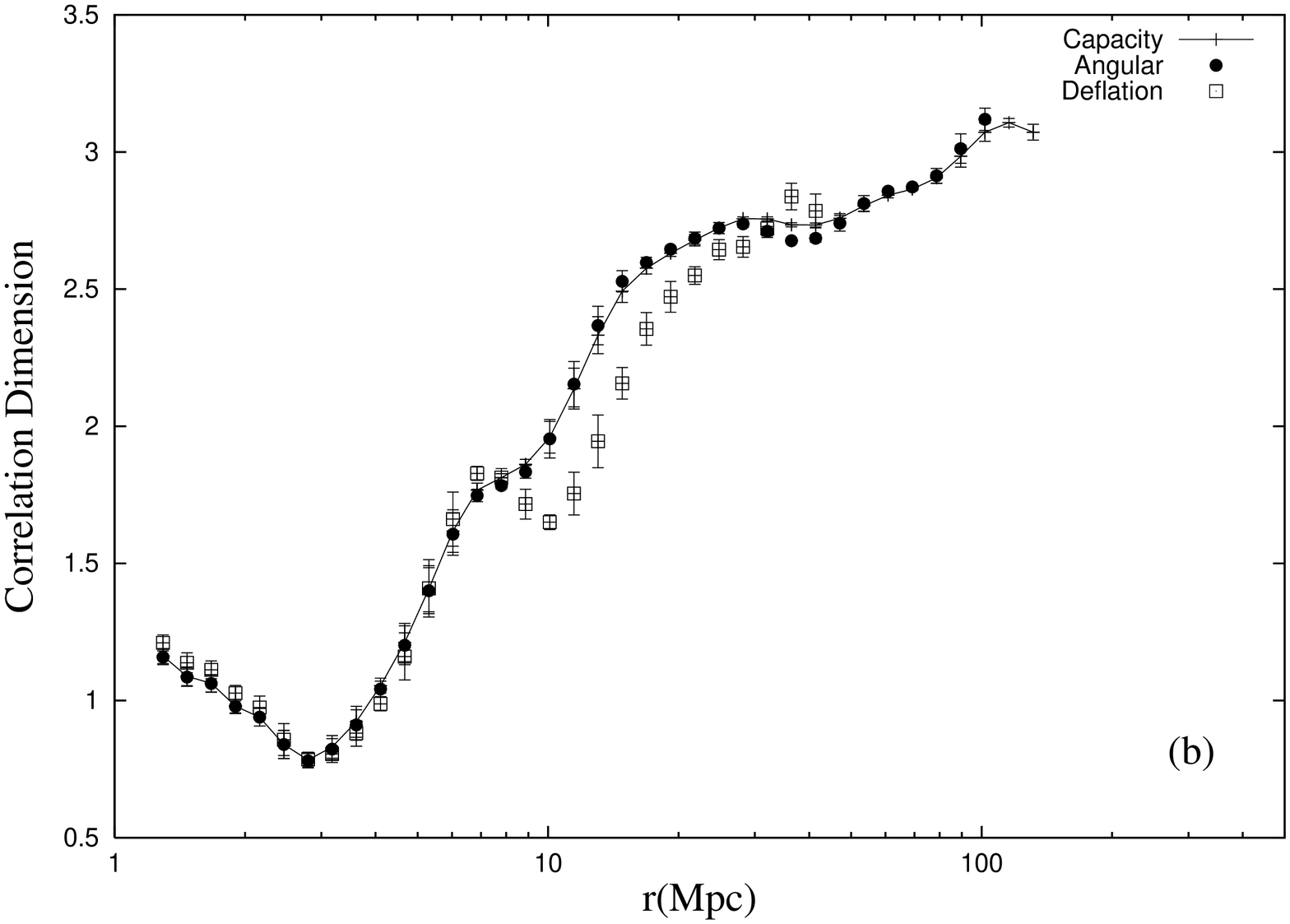,height=8cm,width=7cm,angle=-90}
\end{tabular}
\caption{Fractal analysis of subset of mock PSCz catalogue from
n-body simulation of SCDM model. The sample contains 1835 points.
}
\end{figure*}
\begin{figure*}
\begin{tabular}{cc}
\psfig{file=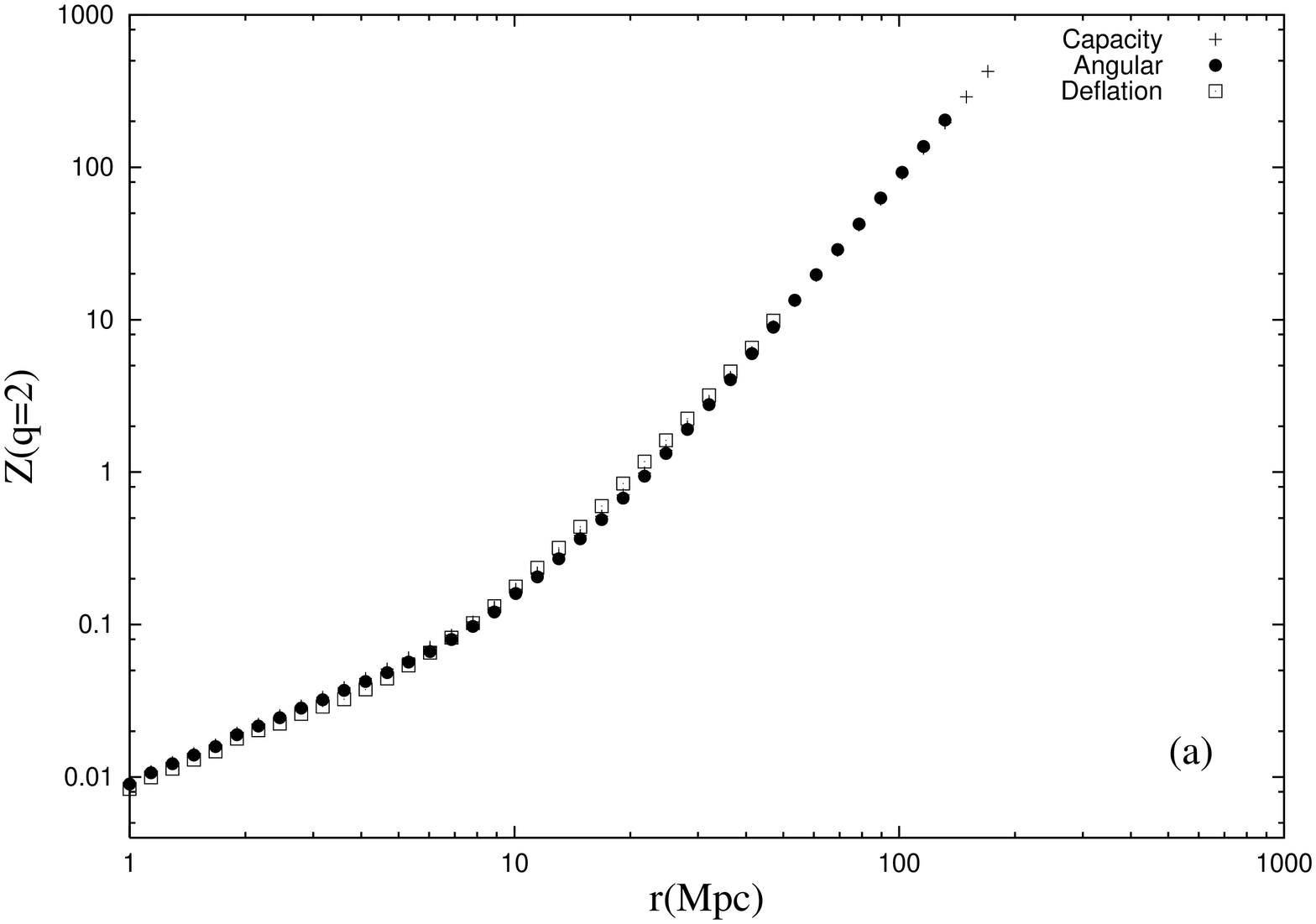,height=8cm,width=7cm,angle=-90}&
\psfig{file=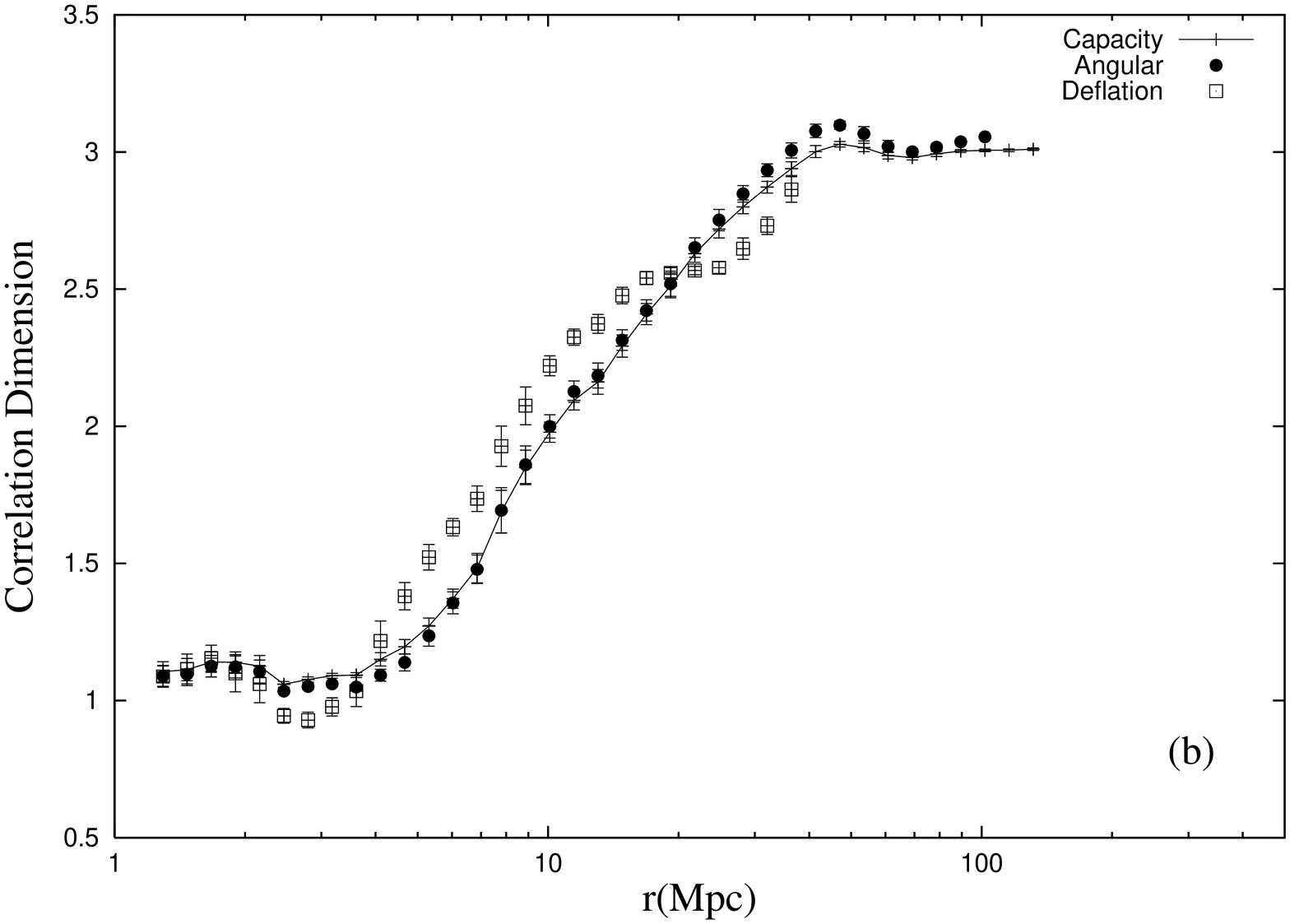,height=8cm,width=7cm,angle=-90}
\end{tabular}
\caption{Results of subset of mock PSCz catalogue from
$\Lambda$CDM model. This set includes 1751 points.  }
\end{figure*}

\begin{table}
\begin{center}
\begin{tabular}{|c|ccc|}\hline
$\Lambda$CDM & Capacity & Angular& Deflation \\ \hline $D_2$
$\begin{array}{c}<5{\rm Mpc} \\ 5\sim 30{\rm Mpc}\\>30{\rm
Mpc}\end{array}$ & $\begin{array}{c}1.11(1)\\2.11(6)\\3.00(1)
\end{array}$ & $\begin{array}{c}1.08(1)\\2.12(6)\\3.04(3)
\end{array}$ & $\begin{array}{c}1.05(2)\\2.28(5)\\2.92(6)
\end{array}$\\ \hline $r_{\rm max}({\rm Mpc})$ &---&$\sim 132$ & $\sim 47$
\\ \hline
\end{tabular}
\end{center}
\caption{Correlation dimensions $D_2$ and for different scales of
the $\Lambda$CDM mock PSCz subsample and $r_{\rm max}$ for angular
correction and deflation methods. Errors are from goodness-of-fit
considerations only.}
\end{table}

\section{Conclusion and Discussion}

We have investigated the treatment of edge effects for the fractal
analysis based on the correlation integral (CI) for various
samples. Generally speaking, on small scales, the CI is almost
unaffected by boundary correction just because there are only a
few points in the vicinity of boundaries. Finite size and
discreteness effects can blur the real dimension on very small
scales, however. The proper scaling law has to be sought on scales
above some critical scale $r_0$ to avoid this problem. Determining
the scale that one needs to reach is not trivial. It has been
argued on the basis of a Voronoi (1908) model that this scale
needs to be about an order of magnitude larger than the Voronoi
tesselation radius $r_v\sim V_v^{1/3}$ in such a model (Sylos
Labini et al. 1996), where $V_v$ is the Voronoi volume and $r_v$
is of order of the mean separation between the Voronoi nuclei.
However, the more relevant criterion for the analysis of very deep
samples like PSCz is that one simply has to have enough galaxies
in the sample to get reliable statistical estimates of the
correlation dimension.

On sufficiently large scales, different boundary corrections lead
to differences in estimates of statistics extracted from the CI.
The capacity correction can introduce extra homogeneity into the
CI, while the deflation method reduces the number of available
points and can consequently generate big fluctuations and lower
statistical significance in the CI. The angular correction leads
to a reasonable compromise between the two former effects,
although it is difficult to apply it to real samples if they have
a complicated geometry, especially if there are holes inside.
However in practice the uncertainties introduced by boundary
effects depend on the properties of samples under analysis. In a
pure fractal set or sample not reaching the homogeneity scale,
capacity correction can fool people into thinking that the
transition to homogeneity has been observed. In samples whose size
extends beyond the  transition scale like the PSCz mock
catalogues, boundary corrections have only a trivial importance in
the analysis. This is a  similar conclusion to that reached by
Lemson \& Sanders (1991) and Provenzale, Guzzo \& Murante (1994),
although they used the conditional density $g(r)$ rather than the
CI which we discuss here.

Whatever the case, it is clear that the capacity correction is
{\em not} the most suitable for exploring the scaling laws for
clustering displayed by galaxy samples.  The angular correction we
propose has less bias and produces less fluctuations than the
others. Of course the relative merits of the different approaches
depend on the precise details of survey shape and sampling
properties. In general one should establish the reliability of a
given result by examining the range of methods. Even the deflation
method, though not optimal, can still be used as auxiliary method
to get an idea of uncertainties or fluctuations.

Our analysis of CfA2-South sample shows that it is indeed fractal,
with dimension $\sim 1.8$ up to $40h^{-1}$ Mpc; there is no sign
of tendency toward a homogeneous distribution in this data set. It
is known that infrared galaxies distribute more homogeneously than
galaxies selected via their luminosity in the optical band.
Infrared galaxies are less likely to be found in the inner parts
of rich clusters, for example. This may account for the
contradiction between the two samples. We can nevertheless expect
that the completion of the next generation of large-scale redshift
surveys, will establish a transition to homogeneity with a scale
somewhere around $30h^{-1}$ Mpc.

The application to subsamples of the PSCz survey indicates that
the features of the distribution of infrared galaxies above scale
$\sim 30 h^{-1}$ Mpc are not modified by any reasonable boundary
correction, which in turn provides further supporting evidence for
a Universe which is homogeneous on large scales. It remains
difficult to put strict error bars on the results, but we can use
the values generated by three treatments to get an idea of the
errors. This constrains the dimension $D_2$ to lie in the range
$2.94\sim 3.06$ on large scales, assuring the validity of the
results in PC. The QDOT survey is one subset of the present PSCz
catalogue, thus the analysis of it by Mart\'{\i}nez \& Coles (1994)
are also supported. Moreover, the observed data behave in
precisely the same manner as the simulation results which are
based on cosmologies in which the Cosmological Principle applies.
However, this satisfying confirmation may not extend to other
samples that have produces claims of large-scale homogeneity. For
example, we have strong reason to suspect that the claimed
tendency-to-homogeneity of the cluster samples from Abell and ACO
catalogues by Borgani \& Mart\'{\i}nez (1994) may not be real, and
the distribution of these clusters remains somewhat uncertain,
owing to the use of the capacity correction in that study. The
difference between angular and capacity dimensions is much smaller
when the conditional density is used.

\section*{Acknowledgment}
The authors wish to acknowledge Dr Spyros Basilakos for providing
mock PSCz catalogues from N-body simulation; we would also like to
thank the anonymous referee for useful criticism and very helpful
suggestions. Jun Pan receives an ORS award from the CVCP.

\end{document}